\newcommand{\CLASSINPUTtoptextmargin}{1cm}
\newcommand{\CLASSINPUTbottomtextmargin}{0.7cm}

\documentclass[journal]{IEEEtran}
\IEEEoverridecommandlockouts                              % This command is only
\usepackage{multirow}
\usepackage[table,xcdraw]{xcolor}
\usepackage{epstopdf}
\usepackage{amsfonts}
\usepackage{amsmath,amssymb}
\usepackage{color,soul}
\usepackage{tikz}
\usetikzlibrary{automata}
\usepackage{color}
\usepackage{verbatim}
\usepackage{comment}
\usepackage{algorithm}
\usepackage[table,xcdraw]{xcolor}
\usepackage{varioref}
\usepackage{gensymb}
\usepackage{comment}
\usepackage{xfrac}
\usepackage{soul}
\usepackage{color}
\usepackage{xcolor}
\usepackage{cite}
\usepackage{ctable}

\usepackage{tikz}
\usetikzlibrary{shapes.geometric, arrows} 
\tikzset{%
    ,io/.style={%
        ,trapezium
        ,trapezium left angle=70,trapezium right angle=110
        ,text width=4cm
        ,minimum height=1cm
        ,text centered
        ,draw=black
        ,fill=blue!10
        }
    ,process/.style={%
        ,rectangle
        ,minimum width=3cm,minimum height=1cm
        ,text width=6cm
        ,text centered
        ,draw=black
        ,fill=blue!10
        }
    ,decision/.style={%
        ,diamond
        ,minimum width=2cm,minimum height=1.5cm
        ,text width=3cm
        ,text centered
        ,draw=black
        ,fill=blue!10
        ,aspect=3
        }
    ,arrow/.style={%
        ,thick
        ,->
        ,>=stealth
        }
    ,line/.style={%
        ,draw
        ,-latex'
        }
    }

\usepackage{caption}
\usepackage{subcaption}
\usepackage{cleveref}
\begin{document}

\title{Ellipsoidal Prediction Regions for Multivariate Uncertainty Characterization}

\author{Faranak~Golestaneh,~\IEEEmembership{Student Member,~IEEE,}
	Pierre~Pinson,~\IEEEmembership{Senior Member,~IEEE,}
         Rasoul~Azizipanah-Abarghooee~\IEEEmembership{Student Member,~IEEE} and~Hoay~Beng~Gooi,~\IEEEmembership{Senior Member,~IEEE}
%\thanks{F. Golestaneh and H. B. Gooi are with the School of Electrical and Electronic Engineering, Nanyang Technological University, Singapore, e-mail: (faranak001@e.ntu.edu.sg).}% <-this % stops a space
%\thanks{P. Pinson is with the Department of Electrical Engineering at the Technical University of Denmark, Denmark, e-mail: (ppin@elektro.dtu.dk).}
%\thanks{Manuscript received September 19, 2014; revised January 11, 2007.}
}
\vspace{-0.8em}
\maketitle

\vspace{-0.3em}
\begin{abstract}
While substantial advances are observed in probabilistic forecasting for power system operation and electricity market applications, most approaches are still developed in a univariate framework. This prevents from informing about the interdependence structure among locations, lead times and variables of interest. Such dependencies are key in a large share of operational problems involving renewable power generation, load and electricity prices for instance. The few methods that account for dependencies translate to sampling scenarios based on given marginals and dependence structures. However, for classes of decision-making problems based on robust, interval chance-constrained optimization, necessary inputs take the form of polyhedra or ellipsoids. Consequently, we propose  a systematic framework to readily generate and evaluate ellipsoidal prediction regions, with predefined probability and minimum volume. A skill score is proposed for quantitative assessment of the quality of prediction ellipsoids. A set of experiments is used to illustrate the discrimination ability of the proposed scoring rule for misspecification of ellipsoidal prediction regions. Application results based on three datasets with wind, PV power and electricity prices, allow us to assess the skill of the resulting ellipsoidal prediction regions, in terms of calibration, sharpness and overall skill.
\end{abstract}
% IEEEtran.cls defaults to using nonbold math in the Abstract.

% Note that keywords are not normally used for peerreview papers.

% For peer review papers, you can put extra information on the cover
% page as needed:
% \ifCLASSOPTIONpeerreview
% \begin{center} \bfseries EDICS Category: 3-BBND \end{center}
% \fi
%
% For peerreview papers, this IEEEtran command inserts a page break and
% creates the second title It will be ignored for other modes.
\IEEEpeerreviewmaketitle

\vspace{-0.6em}
\section{Introduction}
\PARstart{T}{he rapid} deployment and integration of renewable energy generation capacities have increased the level of variability and uncertainty in power systems, possibly also magnified by new electricity consumption patterns. This comes in a context of deregulation of energy markets, eventually resulting in a more complex environment for decision-makers. This calls for the development of a number of forecasting methodologies providing the suitable input to a wealth of decision-making problems in power system operation and electricity market participation, under uncertainty and with a view on risk management~\cite{morales2013integrating}.

The most common deterministic forecasts take the form of single values for each variable of interest, location and lead time. Although these are easier to interpret and to use as input to decision-making problems, they are always subject to errors~\cite{zhang2014review}. Costs induced by such errors often motivate to provide a full picture of potential forecast errors and to accommodate uncertainty estimates in associated optimization problems. Probabilistic forecasting then comprise the appropriate framework to generate that information~\cite{tastu2013short}. Probabilistic forecasts are, however, most often produced in a univariate framework, i.e., still providing uncertainty information for every variable, lead time and location, individually. They are only suboptimal inputs to decision-making when temporal, spatial and/or inter-variable dependencies are to be considered. Besides, due to the inertia in meteorological systems and their impact of renewable power generation, load and electricity markets, such dependencies are expected to be present.

In contrast, a multivariate probabilistic forecast region defines a region where the realization of a multivariate random variable is expected to lie, with a certain probability. A number of optimization methods e.g. stochastic programming are to use scenarios and scenario trees as inputs, which are based on samples from multivariate probabilistic forecasts. However, chance-constrained~\cite{zhang2011chance}, robust~\cite{sousa2011robust, jabr2013adjustable}, interval~\cite{wu2012comparison} optimization require the definition of multivariate probabilistic forecast regions. Only few proposals on generation of multivariate prediction regions can be found in the literature \cite{bessa2015marginal, kolsrud2007time, li2011simultaneous}, referred to as adjusted intervals and Chebyshev-based intervals. The idea is to use already generated sets of scenarios and to deduce prediction regions as a minimum volume that cover given proportions of these scenarios. As reported in~\cite{bessa2015marginal}, the prediction regions with low nominal coverage are too wide and conservative, while difference in the size of the regions for varying nominal coverage rates is low. %Also, the computational cost of the mentioned methods is high and it increases as the number of predicted trajectories increases.

For most optimization problems in power system operation and electricity market applications, multivariate uncertainty sets (another term for multivariate prediction regions when used as inputs to optimization) are assumed to have ellipsoidal geometry~\cite{chassein2016min,bertsimas2004robust}. For example in~\cite{chassein2016min}, two types of cuts are proposed for minmax regret problems with ellipsoidal uncertainty sets, where ellipsoidal uncertainty sets are considered as more flexible and realistic uncertainty sets compared to finite or hyper-boxes. Also in \cite{guan2014uncertainty}, ellipsoidal uncertainty sets are introduced as relevant uncertainty representation in robust unit-commitment.

% Additionally Ref.~\cite{saric2008applications} proposed an approach for ellipsoidal approximation of polyhedral feasibility sets in power systems optimization. Ellipsoid prediction regions can be readily used, or alternatively be approximated by maximum-volume inner polyhedra and minimum-volume outer polyhedra~\cite{barmann2016polyhedral}.

As for univariate probabilistic forecasts, ellipsoid prediction regions ought to provide probabilistically reliable and skillful information about multivariate uncertainty. To the best of our knowledge, there is no established practice so far to generate and evaluate ellipsoidal prediction regions with predefined probability to be used as input to optimization problems. In practice, ellipsoid parameters are chosen based on expert knowledge, assumptions or trial and error. For instance in~\cite{li2015modeling}, a framework is described where the size of the ellipsoids are controlled by a parameter called uncertainty budget which is decided through trial and error, with higher uncertainty budget results in a higher probability and conservativeness. 
%The framework seeks ellipsoids or reasonable size that enclose the most measurements possible.
 A clear disadvantage is that the probability associated with the ellipsoids (their coverage rate) cannot be determined in advance. This is while, in practice, one is most likely interested in having ellipsoid prediction regions with various predefined nominal coverage rates, e.g. 90\%, 95\% or 99\%. 

In this paper, a generic optimization-regression framework is developed to generate the prediction ellipsoids with predefined probability and high performance. The most straightforward assumption about the properties of prediction ellipsoids is to consider them as Gaussian geometries.  In that case, the prediction ellipsoids can be considered as the contours of constant density in multivariate Gaussian distribution where the density is determined by the percentiles of $ \chi^2 $ distribution. However, our empirical investigations revealed that Gaussianity assumption of prediction ellipsoids is not valid for the important random variables in power systems, namely, Photovoltaic (PV) and wind power and electricity price. Those prediction ellipsoids designed based on Gaussianity assumption show very low calibration and reliability. Therefore, in this work, prediction ellipsoids are generated without any restrictive assumption   to skillfully mimic the true underling stochastic process. The proposed prediction ellipsoids are called Ellipsoidal Prediction Regions (EPRs).

In the proposed framework, the centers of  EPRs are point forecasts while the covariance matrix of the ellipsoids are found by either exponential smoothing or Dynamic Conditional-Correlation-GARCH (GARCH-DCC)~\cite{engle2002dynamic}. The choice of covariance matrix forecast technique depends on the inherent uncertainty of random variables. We use exponential smoothing for those random variables with slow-moving covariance matrix while  GARCH-DCC performs much better to forecast a time-varying covariance matrix.  The scale parameters are determined through a optimization procedure using the historical data. A generic objective function is proposed for the optimization phase  leading EPRs to have  lower conservativeness and higher probability guarantees. The proposed model is capable of outputting calibrated EPRs with predefined coverage rates.  

Because the literature on multivariate prediction regions is  at a primitive stage, there is no established evaluation framework for this class of forecasts. A scoring rule is proposed here for quantitative assessment of the prediction ellipsoids based on the essential characteristics required for skilled forecasts, namely reliability (calibration) and sharpness (low conservativeness). A set of empirical experiments are designed to examine the ability of the proposed scoring rule in discriminating possible prediction misspecification in a multivariate context. Additionally, a formulation is proposed to estimate the size  of ellipsoids for bounded random variables. The efficiency of the proposed framework is evaluated for wind and PV power and electricity price. Temporal prediction ellipsoids of dimensions 2, 11 and 24 with the probability of 5\% to 95\% with 5\% increments are generated and evaluated.

%The rest of the paper is organized as follows. In Section II,  the basic definition and formulation of prediction ellipsoids are given. The proposed scoring rule is explained in Section III. The proposed framework to generate  EPRs is described in Section IV. Section V contains the empirical results and finally concluding remarks are given in Section VI.
\vspace{-0.3em}
\section{Ellipsoidal Prediction Regions: Basics and Formulation}\label{npforec}
Let \textbf{X} be a multivariate random variable of dimension $D$. In case of temporal dependency, \textbf{X} can be described  as $\textbf{X}_t=\{X_{t+k_1},...,X_{t+k_D}\}$ with $k_i \:\forall i$ as the forecast horizons. To simplify the notation, hereafter $ \textbf{X}_t $ is denoted as $\textbf{X}_t=\{X_{t+1},...,X_{t+D}\}$.

Let $PE^{\alpha}$ be the prediction ellipsoid with the nominal coverage rate as $ \alpha $, where $ \alpha $ represents the ratio of realizations of $ \textbf{X} $ inside $PE^{\alpha}$~\cite{pope2008algorithms}.
  \begin{equation}
  \label{Eq:EllipsosoidTypical}
PE^{\alpha}:  (\textbf{x}-\mu)^\top \varSigma^{-1}(\textbf{x}-\mu)\leq \varUpsilon^{\alpha}\\
  \end{equation}
  where $ \mu= E(\textbf{X})$ is the mean vector of the random variable.   $ \varSigma= E[(\textbf{X}-\mu)(\textbf{X}-\mu)^\top] $ is the covariance matrix. $ \varUpsilon^{\alpha} $ is called the scale or robust parameter. It should be noted that hereafter, upper case letters symbolize random variables while lower case letters express their realizations.

 When $ \textbf{X}$ follows a multivariate Gaussian distribution,  $ \textbf{X}\sim\mathcal{MVN}(\mu,\Sigma) $, $ \textbf{x} $ in \eqref{Eq:EllipsosoidTypical} describe contours of constant density for the $D$-dimensional normal distribution. In this case, the scale parameters are the percentiles of  $ \chi^{2} $ distribution as
  \begin{equation}
  \label{Eq:Chi-sq_Main}
  PE^{\alpha}: (\textbf{x}-\mu)^\top \varSigma^{-1}(\textbf{x}-\mu)\leq \chi^{2}_{D}(\alpha)\\
  \end{equation}
with $ \chi^{2}_{D}(\alpha) $ as the lower 100\textsuperscript{th} percentile of $ \chi^{2} $ with $ D $ degrees of freedom, the ellipsoid in \eqref{Eq:Chi-sq_Main} has probability of $ \alpha $.

In robust optimization, the scale parameter is also called the uncertainty budget and it controls the trade-off between robustness and performance. The value of uncertainty budget usually  is selected arbitrarily or by trial and error in the range of $ [0,D^2]$~\cite{li2015modeling}. Let us call the ellipsoids characterized by the uncertainty budget as the robust ellipsoids.
\vspace{-0.3em}
\section{Evaluating the Skill of Ellipsoidal Prediction Regions}

The aim of designing a scoring rule is to provide a theoretically principled framework for quantitative assessment of predictive performance of ellipsoidal prediction regions. In general, two properties are required for  probabilistic forecasts, namely calibration and sharpness. In the context of  ellipsoidal regions  similar to the case of univariate quantiles, calibration is referred to the proximity of the nominal coverage rate of an ellipsoid to its observed coverage. The definition of the sharpness though is more challenging in this new context. One can consider the volume of ellipsoids as the most straightforward representation of sharpness.  Here, a scoring rule is proposed for verification of predictive performance of EPRs.
\vspace{-0.3em}
 \subsection{Formulation}
   The forecaster is always looking for reliable and calibrated prediction regions with a minimal area or volume possible to reduce the conservativeness. Sharpness and calibration can be assessed simultaneously through a  skill score. A negatively-oriented skill score is expected to assign  the lowest score value to the actual (true) ellipsoid. The proposed ellipsoidal skill score is given by
\begin{equation}
\label{skillscore1}
%\vspace{-0.3em}
Sc_{\alpha_i}=|\frac{1}{T}\sum_{t=1}^{T} { \left( \xi_{t}^{\alpha_i}-\alpha_i \right) {(V_t^{\alpha_i})}^{\dfrac{1}{D}}}\rvert\\
\end{equation}
where  $ T $  is the number of multivariate ellipsoids available.  $\xi_{t}^{\alpha_i}$ is an indication variable which is equal to 1 if the observed trajectory is inside the predicted geometrical region and is 0  otherwise. The observed trajectory is inside the ellipsoid if it   satisfies \eqref{Eq:EllipsosoidTypical}. $\alpha_i$ shows the nominal coverage rate of the  predicted geometrical region.  $V_t^{\alpha_i}$ is the volume of multivariate  ellipsoid with nominal coverage rate $ \alpha_i $ at time $t$ and  it  is calculated by  
\begin{equation}
\label{Vol_ellipsoid}
V_t^{\alpha}= \dfrac{\pi^{\frac{D}{2}}}{\Gamma(\frac{D}{2}+1)} \sqrt{ (\varUpsilon_{t}^{\alpha}) ^D det (\varSigma_t )}\\
\end{equation}
\lowercase{where} $ \Gamma $ represents Gamma function~\cite{li2015modeling}.

To get a single score for all prediction ellipsoids with nominal coverage rates $ \alpha_i , i=1,...,m$, one can sum individual scores as
\begin{equation}
	\label{skillscoreAllm}
%	\vspace{-0.3em}
	Sc=\sum_{i=1}^{m} { \left( Sc_{\alpha_i} \right) }
	\vspace{-0.2em}
\end{equation}

In order to assess calibration only, as for univariate probabilistic forecasts, one can calculate the observed coverage rate and compare it with nominal one. The observed coverage rate can be calculated as
\begin{equation}
	\label{calib}
%	\vspace{-0.8em}
	\hat{\alpha}_i = \frac{1}{T} \sum_{t=1}^{T} {\xi_{t}^{\alpha_i}}
	\vspace{-0.7em}
\end{equation}
\vspace{-0.4em}

It is to be noted that the formulation given in \eqref{Vol_ellipsoid} is accurate if the ellipsoids do not exceed the feasible limits of random variables. The prediction ellipsoids for PV and wind power are bounded between zero and nominal capacity of the corresponding wind or PV installation. Therefore, the feasible volume of each \textit{D}-dimensional  prediction ellipsoid is the intersection of that ellipsoid and a \textit{D}-dimensional polyhedron. Calculation of the volume of the intersection analytically is intractable. However, one can use a Monte Carlo based approach to estimate the feasible volume numerically~\cite{li2015modeling}. The proposed methodology to estimate the feasible volume of EPRs is explained in Appendix \ref{Section:Volume}.
\vspace{-0.4em}	
\subsection{Evaluation of discriminating capability}
The possible prediction errors in ellipsoidal context are the errors in prediction of the center, the correlation (covariance) matrix, variance in each dimension and the scale parameter. To investigate the ability of the skill score proposed in \eqref{skillscore1} to detect possible prediction errors, the following experiments are designed. In all experiments $T=10,000$  vectors of realizations of random variable $ \textbf{X} $ are generated from the actual Gaussian  density. Let $ \textbf{x}_t $ be the realization of  $ \textbf{X} $ at time $ t $ with $ x_{i} $ as  its element at the $ i^{th} $ dimension. Let the actual density be defined with zero mean and unit variance of dimension $ D=24 $, and covariance function as
 \begin{equation}
 \label{TrueCov}
 \vspace{-0.3em}
 \varSigma(x_i,x_j)=\sigma_i \sigma_j \exp (-\dfrac{\lvert i-j \rvert}{4}) \quad i,j=1,...,D
 \end{equation}
with $ \sigma_i \: \forall i $ as the variance of $ \textbf{X} $ in its $ i^{th} $ dimension.
\begin{enumerate}
\item \textit{Misspecified mean} (\textit{center}):
 In this scenario, the prediction ellipsoids are assumed to have the correct covariance matrix as described in \eqref{TrueCov} and the correct scale parameters as given in \eqref{Eq:Chi-sq_Main} but erroneous  center (mean) as $ \hat{\mu}_i=\Xi(-1,1), i=1,...,D  $.  $ \Xi(a,b) $ is a function which generates decimal values between $ a $ and $ b $  from the Uniform distribution. For 10,000 successive times, 10,000 mean vectors are generated and assumed to be the center of 10,000 prediction ellipsoids.
 
 \item \textit{Misspecified variance}: The prediction ellipsoids for this case are formulated with true mean and scale parameter but with wrong variance  as $ \hat{\sigma}_i=\sigma_i+\Xi(-0.15,1), i=1,...,D  $.

 \item \textit{Misspecified covariance model and strength}: The prediction ellipsoids are modeled with the actual center, variance and scale parameters but with misspecified correlation models and  correlation strengths as
 \begin{equation}
 \label{CovM}
 \vspace{-0.3em}
 \hat{\varSigma}(x_i,x_j)=\sigma_i \sigma_j  (1+\dfrac{\lvert i-j \rvert}{r})^{-1} \quad i,j=1,...,D
 \end{equation}
 with $ r= \Xi(2,6)$.
 
  \item \textit{Misspecified scale parameter}: Prediction ellipsoids in this case have the actual mean, variance, covariance matrix but they are characterized with wrong  scale parameters $ \varUpsilon_{t}^{\alpha} $.  
 \begin{equation}
 \varUpsilon_{t}^{\alpha}=\Xi(0.01\chi^{2}_{D}(\alpha),3 \chi^{2}_{D}(\alpha))
 \end{equation}
 Subject to: if $ \alpha_i>\alpha_j $, then  $ \varUpsilon_{t}^{\alpha_i} >  \varUpsilon_{t}^{\alpha_j}$ 
\end{enumerate}
\begin{figure}[t]
	\vspace{-0.4em}
	%\begin{center}
	\centering
	\includegraphics[width=8cm,height=7cm]{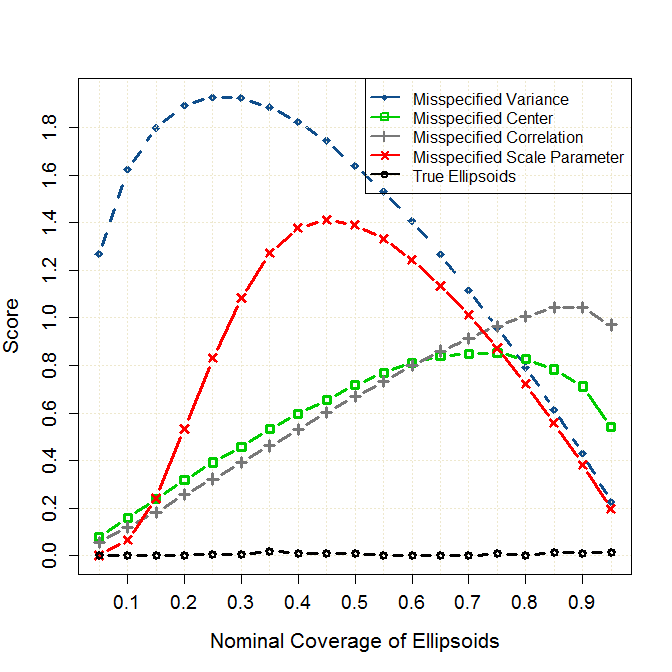}
	\caption{{Skill scores for true Gaussian ellipsoids versus those for various misspecified ellipsoids.}}
	\label{Fig:CheckScore}
	%\end{center}
	\vspace{-0.4em}
\end{figure}
Fig. \ref{Fig:CheckScore} demonstrates the discrimination ability of the skill score introduced in \eqref{skillscore1} in detecting the various types of misspecification. The  scores are calculated per $\alpha$ ranging from 0.05 to 0.95 with 0.05 increments. As one can see,  the best scores are obtained for the true ellipsoids. The relatively lower scores of true ellipsoids with respect to the other four misspecified  ellipsoids allow for relying on the proposed skill score to discriminate a  good ellipsoidal representation of uncertainty against an incorrectly specified one.
\vspace{-0.2em}
\section{Ellipsoidal Prediction  Regions}

%The aim of this work is to develop a framework for generation of ellipsoidal prediction regions for random variables in power systems.

 %To the best of our knowledge, there has not been any established methodology so far to produce predictive ellipsoids with predefined coverage rates. 
 
 The proposed EPRs are constructed through an optimization-regression framework. The optimization phase is conducted only once for  historical data to find the scale parameters. The first regression phase deals with univariate point forecasting. The second regression phase updates the covariance matrix of point forecast errors using the forecast errors calculated up to time $t$. The goal is to generate  $ m $  EPRs with nominal coverage rates $ {A}=[\alpha_1 \: \alpha_2  \: ... \:\alpha_m] $.
\subsubsection{Parameters specifications}
The EPRs are given by
\begin{equation}
\label{elipFormula}
	(\textbf{x}_t-\mu_t)^\top \varSigma^{-1}_{t}(\textbf{x}_t-\mu_t)\leq \varUpsilon^{\alpha}\\
\end{equation}

$ \mu_t $:  Considered to be the point forecasts for time $ t $. Denote $\hat{\textbf{x}}_t=\{\hat{x}_{1,t},\hat{x}_{2,t},...,\hat{x}_{D,t}\}	 $, with $ \hat{x}_{i,t}, \: \forall i $  as the point forecast for time $ t $ and dimension $ i $ where  $ \hat{x}_{i,t} $ for each dimension is generated independently. Let us call $ \hat{\textbf{x}} $ as the predicted or estimated trajectory and $ \textbf{x} $ as the measured or true trajectory.    With this definition, $ \mu_t $ is considered to be equal to $\hat{\textbf{x}}_t$. 

$ \varSigma $: The covariance matrix of point forecast errors. Covariance matrix is a critical input in multivariate dependency characterization. To generate skilled predictive ellipsoids, if correlations between the random variables of interest and/or their variances change over time, forecasting the future covariance/correlations is required. 

Here, three established methods are suggested for covariance matrix forecasting. These methods are rolling historical correlations, exponential smoothing~\cite{zakamulin2015test} and DCC-GARCH. The first two methods have been widely used in literature and practice because of their simplicity. However, they are not able to provide reliable estimates of correlations unless the covariance/correlation is either constant over short period of time or varies very slowly over time~\cite{zakamulin2015test}. On the other hand, in econometric literature, DCC-GARCH is reported to be capable of estimating time-varying covariance matrix. DCC-GARCH is a computationally efficient multivariate model, which has the flexibility of the univariate GARCH models while it parameterizes the conditional correlations directly. The reader is referred to~\cite{engle2002dynamic,engle2001theoretical}  for the formulation of the method. 

 In this work, the covariance matrix is updated  any time a set of  measurements/point forecasts for \textbf{X} is available.   

$ \varUpsilon^{\alpha}$: To be obtained through a optimization framework. Denote $\mathbf{\Upsilon}$ as the compact form of $ \varUpsilon^{\alpha} $
as
$\mathbf{\Upsilon}=[\varUpsilon^{\alpha_1} \: \varUpsilon^{\alpha_2} \: ... \: \varUpsilon^{\alpha_m}]$.

%An advantage of  using point forecasts in the proposed methodology is that it allows to benefit from the developments and improvements in point forecasting literature, in order to quantify the uncertainty associated with forecasts in a multivariate context.

\subsubsection{Utility function  and constraints} 	
The idea is to optimize $\mathbf{\Upsilon}$ such that the EPRs present desired probability guarantees and conservativeness. This implies that observed coverage rates of the predictive geometrical regions should be as close as possible to the nominal coverage rates while their volumes are kept minimal. Therefore, the potential objective functions for optimization can be introduced as 
\begin{equation}
\label{Eq:objective}
%\vspace{-0.3em}
 \arg_{\mathbf{\Upsilon}} \min \frac{1}{m}\sum_{i=1}^{m}(|\frac{1}{T}\sum_{t=1}^{T} { \left( \xi_{t}^{\alpha_i}-\alpha_i \right) {(V_t^{\alpha_i})}^{\dfrac{1}{D}}}\rvert)
\end{equation}
subject to	
\begin{equation}
\label{Eq:crossing}
\text{ if  }  \alpha_i>\alpha_j , \text{then   }   \varUpsilon_{t}^{\alpha_i} >  \varUpsilon_{t}^{\alpha_j} 
\end{equation}
%\begin{equation}
%\label{TehtaLimit}
%0.2 \chi^{2}_{D}(\alpha_i) \leq \varUpsilon^{\alpha_i} \leq 4 \chi^{2}_{D}(\alpha_i) \quad \forall i\\
%\end{equation}
where $ \lvert.\lvert $ is the absolute value function and $ T $  is the number of measurements in the training set.  $V_t^{\alpha_i}$ is the volume of multivariate  ellipsoid with nominal coverage rate $ \alpha_i $ at time $t$. The constraint in \eqref{Eq:crossing} is considered in order to avoid crossing EPRs.

%Theoretically the allowed range of the $\varUpsilon^{\alpha_i}$ can be 0 to infinity. However, this range can be reduced  to facilitate optimization.  We have set the search space for $\mathbf{\Upsilon}$ as in \eqref{TehtaLimit}.
\subsubsection{Using the optimized  $ \mathbf{\Upsilon} $ to generate EPRs}
 The optimization process is conducted only once for each stochastic process  using the training data. Then, to generate prediction ellipsoids for each time $ t>T $, the point forecast trajectory is generated and is used as the centre of the ellipsoids. Covariance matrix is updated and then by having the optimized $ \mathbf{\Upsilon} $ as the scale parameters, EPRs are readily available.
\vspace{-0.5em}
\section{Results}
In this section, the applicability of the proposed method   for generation of skilled  EPRs is investigated.
\vspace{-0.5em}
\subsection{Data} \label{Data}
As the basis for investigation of  EPRs,  three datasets  are used here.  The datasets include data for electricity price, wind  and PV power. The datasets have been prepared for the Global Energy Forecasting Competition (GEFCom) 2014 and are available online~\cite{website55}. For all three datasets, the resolution of data is of one hour and forecast horizons are 1- to 24-hour ahead. The datasets are briefly described below. For the full specifications, the reader is referred to~\cite{hong2016probabilistic}.

\begin{itemize}
\item   \textbf{Price data:}{ This dataset includes zonal load (MW) and  system load forecasts (MW) as the predictors and locational marginal price (\$/MW) as the predictand.  The dataset covers about three years worth   data (from January 1st, 2011 to December 17th, 2013). The available data is divided into two   parts including 550 and 532 days worth of data as the training and the evaluation sets, respectively. Price values are normalized by the maximum price available in the data.}

\item  {\textbf{Wind power data:} The wind data provides wind power output series from 10 wind farms in Australia. The data for the second wind farm is used in this paper. The predictors are zonal and meridional wind components  forecasts at two heights, 10 and 100 m above ground level, generated by  the European Centre for Medium-range Weather Forecasts (ECMWF).  The predictand is wind power generation. The predictions were issued every day at midnight.   The period for which both predictions and  measurements are available is from January 2012 to December 2013. Training data covers the period from January 2012 to April 2013. The data from May 2013 to December 2013 is used for skill verifications. Power measurements are normalized by the nominal capacity of the corresponding wind farm.}
\item  \textbf{PV power data:}  Explanatory variables include 12  independent  variables as the output of Numerical Weather Prediction (NWP) provided by ECMWF and the predictand is PV power generation. Data for the period of April 2012 to the end of June 2014 for three zones is available.  The date for the first zone is used here. Training data covers the period from April 2012 to the end of May 2013. The data from June 2013 to the end of January 2014 is used for skill evaluation.  Power measurements are normalized by the nominal capacity of the corresponding PV installation.
\end{itemize}
\vspace{-0.4em}
\subsection{Set-up}
 For wind power and electricity price, the temporal correlations of 1- to 24-hour ahead prediction are studied. For PV power data, the temporal dependency of hourly PV generation from 7 am to 5 pm are taken into account.  The BOBYQA algorithm is deployed as the global search engine and the Generalized Simulated Annealing (GenSA) is used for local search.  The function ``optimx'' and ``GenSA'' in R are used for optimization. BOBYQA is selected  because it gives the most optimal solutions for the problem in this study comparing to the other solvers available in ``optimx''. The window size $ w $ in rolling historical correlations and exponential smoothing methods is considered to be 500 for all simulations. The  decay constant in exponential smoothing method is chosen to be 0.99.  
%The value of $ \lambda $ is considered to be equal to $ \nu $.
 All the analyses below are conducted based on the results obtained for the evaluation data.

The point forecasting and covariance matrix foretasting set-ups  are as follows:

\textit{PV Power}: Data is preprocessed as explained in~\cite{huang2016semi}. The $k$-nearest neighbors (KNN) algorithm and Support Vector Machines (SVM) are deployed to provide forecasts. In order to predict PV power for a particular time, first, the $k$
nearest neighbors of the explanatory variables available for that time are found.
Then, those neighbors are considered as the training set to formulate the
forecasting model. In other words, for each particular time, a new forecasting model
is trained using the $k$ nearest neighbors found within the historical data. $k$ in KNN is considered to be equal to
300. The cost value and the gamma parameters in SVM are found by 5-fold cross-validation. 

DCC-GARCH(1,1) rolling forecast is used to predict the time-varying covariance matrix of PV power because it found to be more efficient than rolling historical correlations and exponential smoothing methods. A moving window of size 300  is used in the rolling estimation. An ARMA(0,1)-GARCH(1,1) is used as the univariate estimator for  the conditional mean and variance in the DCC model.  

\textit{Wind Power}: Forecasts are provided by SVM because it is found to be  more efficient than the combination of KNN and SVM. Since all wind farms
are adjacent to each other, the NWPs available for all ten wind farms are used as the explanatory variables to generate forecasts for farm 2.

The set-up for covariance matrix forecasting is similar to PV power case.

\textit{Electricty Price}: Generalized linear regression is chosen  because it shows better performance than SVM, ELM~\cite{golestaneh2016very} and KNN for the price data. We found the price  values less volatile comparing to PV and wind power. The covariance matrix of price data varies very slowly in time. Therefore, here the exponential smoothing method is found to be  more efficient in forecasting the covariance matrix. 

The point forecast accuracy in terms of Root Mean Score Error (RMSE) for all three datasets are given in Table \ref{Table:RMSE}.

\vspace{-0.4em}
\subsection{EPRs visualization}
\begin{table}[t!]
	\centering
	\caption{Point forecasts accuracy in percent form (\%)}
	\label{Table:RMSE}
	\begin{tabular}{c|cc}
		\specialrule{1.5pt}{0pt}{0pt}
		& \textbf{\begin{tabular}[c]{@{}c@{}}RMSE \\  (Train Data)\end{tabular}} & \textbf{\begin{tabular}[c]{@{}c@{}}RMSE\\  (Test Data)\end{tabular}} \\ \specialrule{1pt}{0pt}{0pt}
		\textbf{Price}      & 3.25                                                                   & 3.8                                                                  \\ \hline
		\textbf{Wind Power} & 16.83                                                                  & 11.82                                                                \\ \hline
		\textbf{PV Power}   & 8.73                                                                   & 10.68                                                                \\ \specialrule{1.5pt}{0pt}{0pt}
	\end{tabular}
	\vspace{-0.7em}
\end{table}
\begin{figure*}[t!]
	\begin{tabular}[c]{ccc}
		\begin{subfigure}{.33\textwidth}
			\centering
			\includegraphics[height=5.5cm,width=\linewidth]{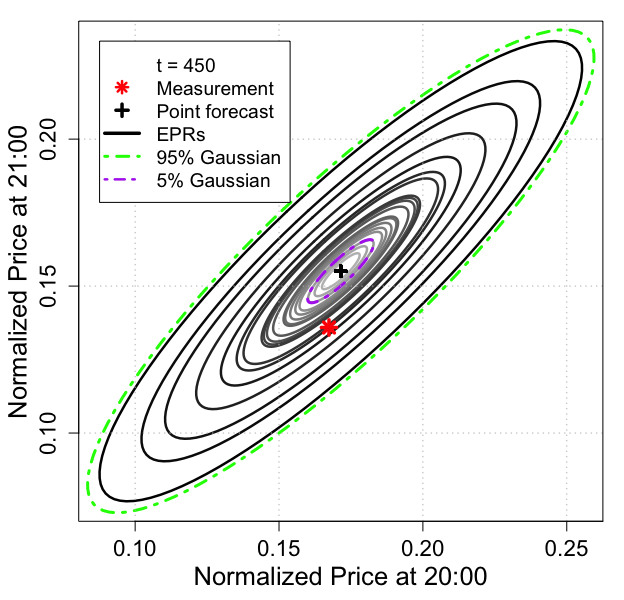}
			%  \caption{Price: Prediction ellipsoids of dimension 24}
			\label{fig:OPESPrice}
		\end{subfigure}
		\begin{subfigure}{.33\textwidth}
			\centering
			\includegraphics[height=5.5cm,width=\linewidth]{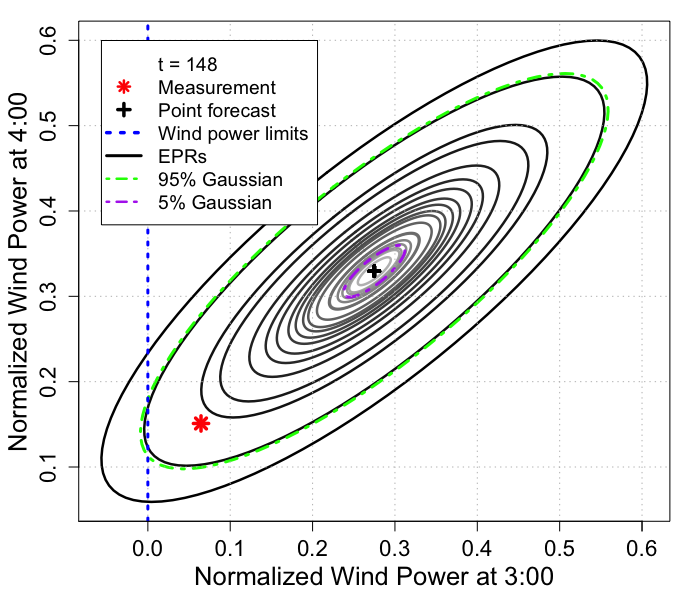}
			% \caption{Wind Power: Prediction ellipsoids of dimension 24}
			\label{fig:OPEsWind}
		\end{subfigure}
		\begin{subfigure}{.33\textwidth}
			\centering
			\includegraphics[height=5.5cm,width=\linewidth]{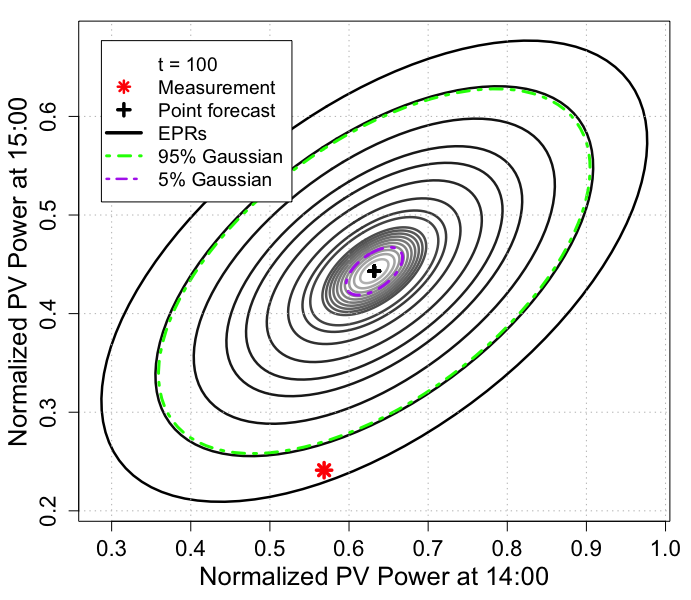}
			% \caption{PV Power: Prediction ellipsoids of dimension 11}
			\label{fig:OPEsPV}
		\end{subfigure}
	\end{tabular}
	\caption{19 Optimal EPRs with probabilities ranging from 0.05 to 0.95 by 0.05 increments (from the lightest to the darkest), for three randomly selected days from the evaluation data of  electricity price, wind and PV power. Character $ t $ denotes the day number in the evaluation datasets. \label{Fig:19OPEs}}
	\vspace{-0.5em}
\end{figure*}
\begin{figure*}[!t]
	\vspace{-0.4em}
	%\begin{center}
	\centering
	\includegraphics[width=19cm,height=5.5cm]{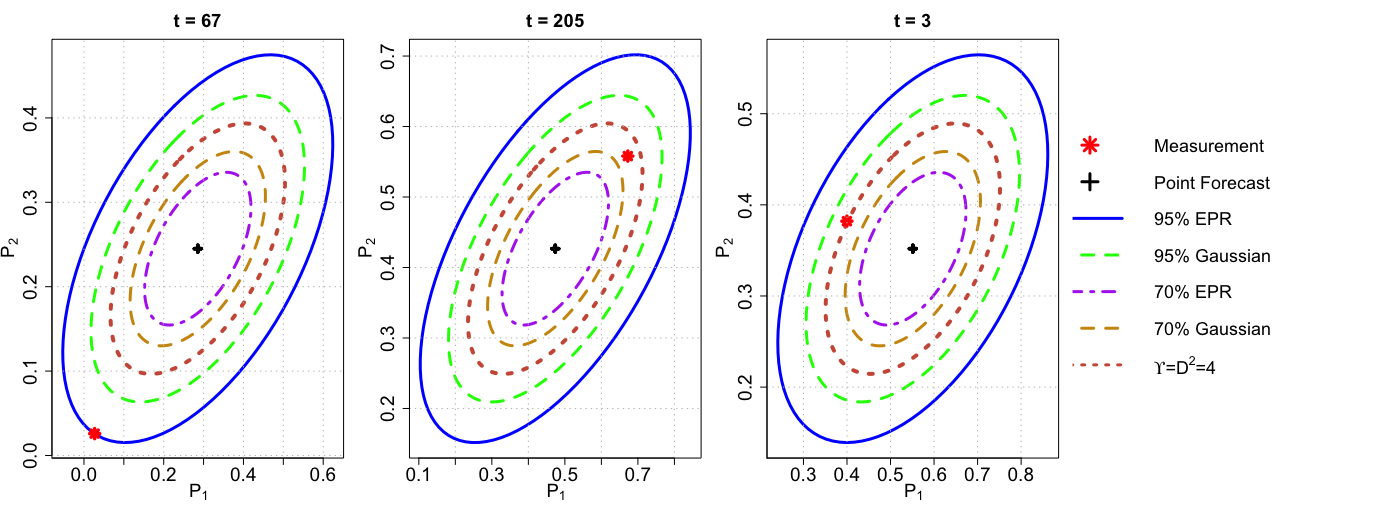}
	\caption{{PV power: Prediction ellipsoids of dimension two for three days from the evaluation dataset. $ \text{P}_1 $ and $ \text{P}_2 $ represent normalized predicted PV power for 14:00  and 15:00, respectively. Character $ t $ denotes the day number.}}
	\label{Fig:PEComparePV}
	%\end{center}
	\vspace{-0.4em}
\end{figure*}
\begin{figure*}[!t]
	\vspace{-0.4em}
	%\begin{center}
	\centering
	\includegraphics[width=19cm,height=5.5cm]{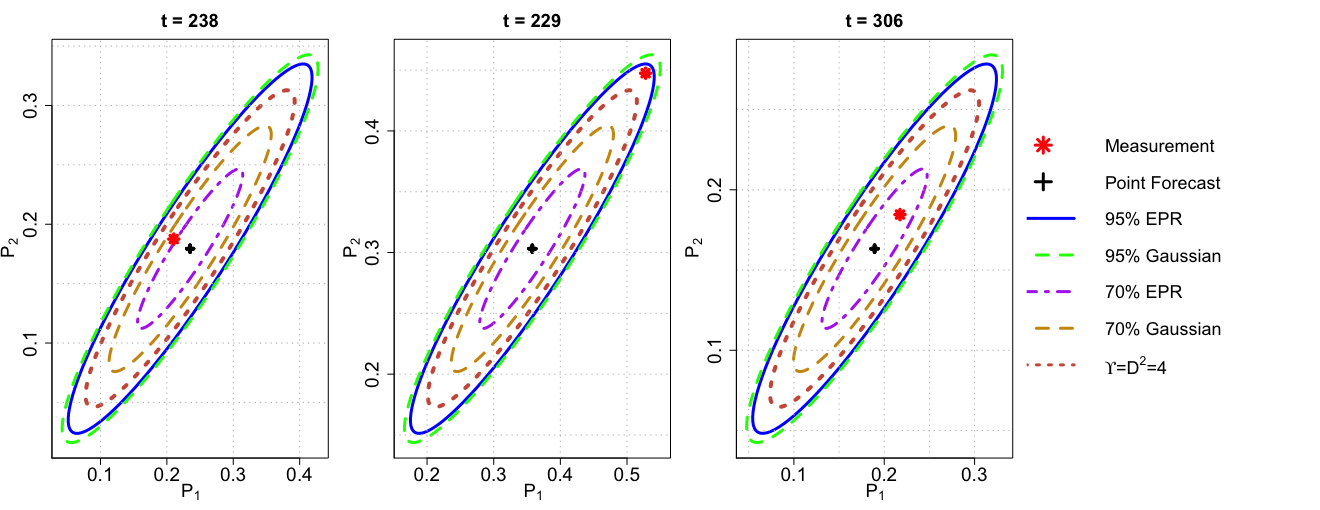}
	\caption{{Price: Prediction ellipsoids of dimension two for three randomly selected days from the evaluation dataset. $ \text{P}_1 $ and $ \text{P}_2 $ represent normalized predicted price of energy for 20:00 and 21:00, respectively. Character $ t $ denotes the day number.}}
	\label{Fig:PEComparPrice2}
	%\end{center}
	\vspace{-0.4em}
\end{figure*}
In order to visualize the EPRs, two dimensional prediction ellipsoids are generated for all three datasets. For the electricity price data, prediction ellipsoids describing the joint uncertainty of the price at 8:00 pm and 9:00 pm are generated. For  wind  power, the bivariate uncertainty sets are obtained for 3:00 am and 4:00 am while those of PV power data are generated for 2:00 pm and 3:00 pm.

In Fig. \ref{Fig:19OPEs}, 19  EPRs   with probabilities ranging from 0.05 to 0.95 by 0.05 increments, for three   randomly selected days are illustrated. One can notice that as the nominal coverage rates of the ellipsoids increase, the EPRs become larger.  The blue dotted line in Fig. \ref{Fig:19OPEs} for wind power describes the actual generation limits,  bounded between 0  and 1 pu. When using prediction ellipsoids as constraints in interval or robust optimization, these practical  limits should be added  to the optimization framework as the additional constraints. 
\begin{figure*}[t!]
	\begin{tabular}[c]{ccc}
		\begin{subfigure}{.33\textwidth}
			\centering
			\includegraphics[width=\linewidth,height=5.5cm]{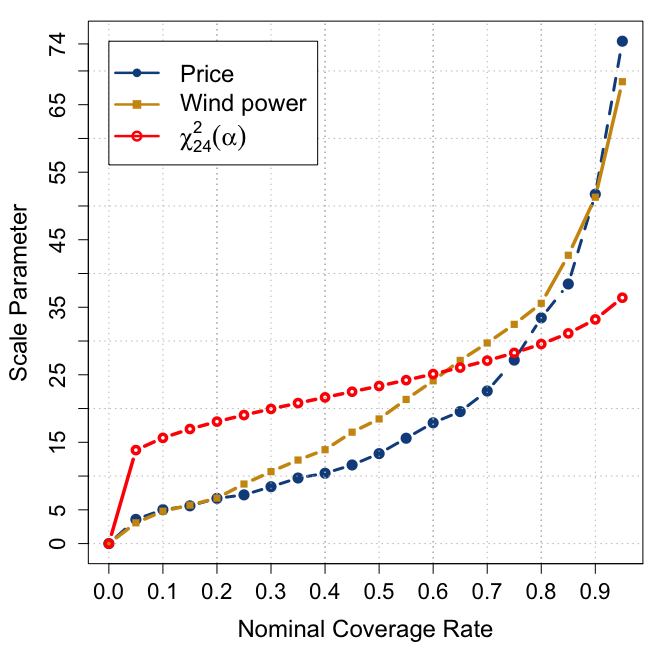}
			\caption{Dimension 24}
			\label{fig:gamma24}
		\end{subfigure}
		\begin{subfigure}{.33\textwidth}
			\centering
			\includegraphics[width=\linewidth,height=5.5cm]{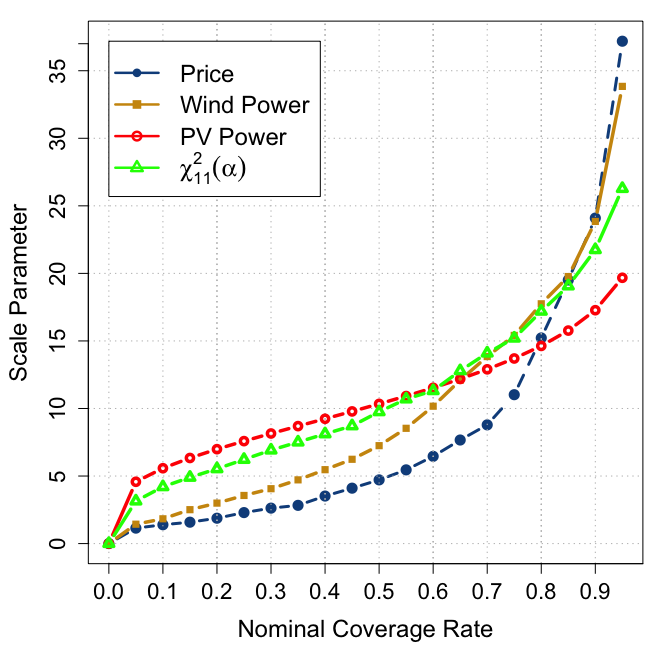}
			\caption{Dimension 11}
			\label{fig:gamma11}
		\end{subfigure}
		\begin{subfigure}{.33\textwidth}
			\centering
			\includegraphics[width=\linewidth,height=5.5cm]{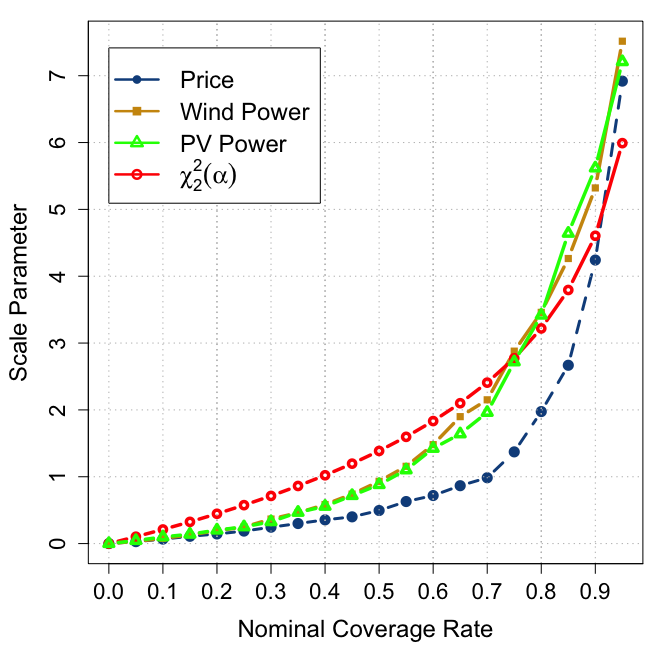}
			\caption{Dimension 2}
			\label{fig:gamma2}
		\end{subfigure}%
	\end{tabular}
	\caption{The scale parameters for the prediction ellipsoids with nominal coverage rates ranging from 0.05 to 0.95 by 0.05 increments. \label{fig:gamma}}
	\vspace{-0.5em}
\end{figure*}
\begin{figure*}[t!]
	\begin{tabular}[c]{ccc}
		\begin{subfigure}{.33\textwidth}
			\centering
			\includegraphics[width=\linewidth,height=5.5cm]{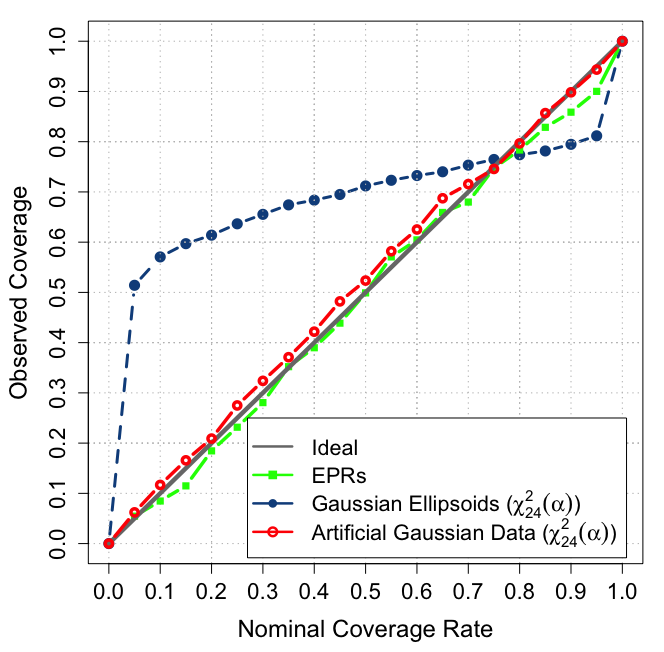}
			\caption{Price, dimension 24}
			\label{fig:QQPrice}
		\end{subfigure}
		\begin{subfigure}{.33\textwidth}
			\centering
			\includegraphics[width=\linewidth,height=5.5cm]{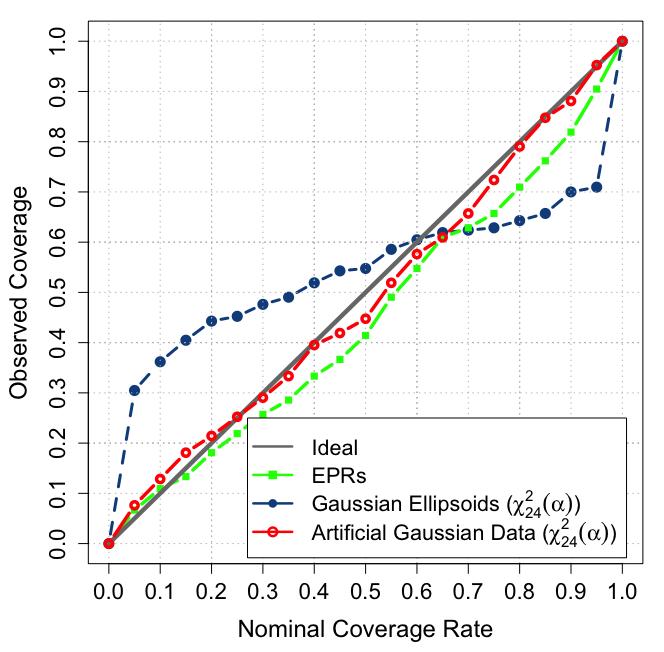}
			\caption{Wind Power, dimension 24}
			\label{fig:QQWind}
		\end{subfigure}
		\begin{subfigure}{.33\textwidth}
			\centering
			\includegraphics[width=\linewidth,height=5.5cm]{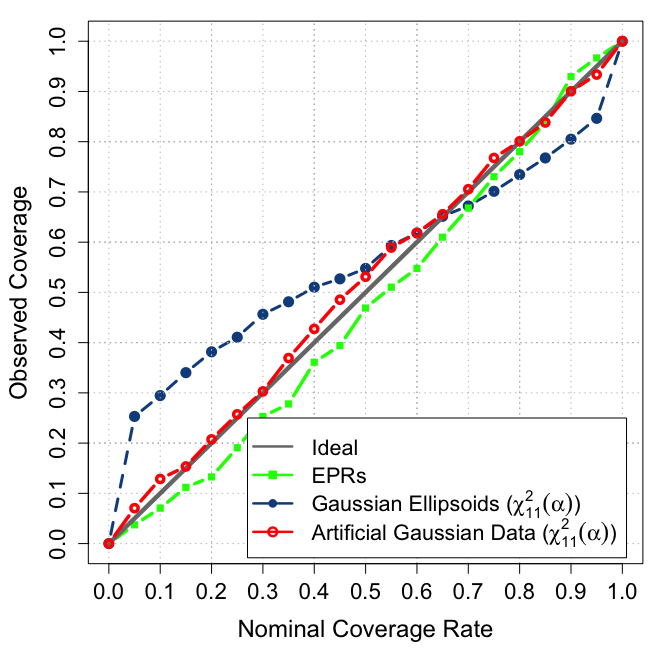}
			\caption{PV Power, dimension 11}
			\label{fig:QQPV}
			\vspace{-0.6em}
		\end{subfigure}%
	\end{tabular}
	\caption{Calibration of the prediction ellipsoids with nominal coverage rates ranging from 0.05 to 0.95 by 0.05 increments. \label{fig:QQ}}
	\vspace{-0.5em}
\end{figure*}
%Comparing the EPRs  for price versus wind and power data demonstrates the higher variability and uncertainty levels in wind and power generations. The standard deviations ($ \sigma $) for price measurements and forecasts at 6:00 pm to 7:00 pm are only 9.27\% and 2.23\% respectively while those of the wind power at 6:00 am to 7:00 am are 28.83\% and 17.68\%. Similarly the standard deviations of PV generations and PV power forecasts at 10:00 to 11:00 am are 24.01\% and 15.38\%. This implies that with the assumption of Gaussian marginal distributions, the 95\% confidence intervals ($ 2\sigma $ around the mean) for the mean of the wind and PV generations are wide.

 As is shown in Fig. \ref{Fig:19OPEs}, although 95\% Gaussian ellipsoids for PV and wind power are smaller than the 95\% EPRs, 5\% Gaussian ellipsoids are larger than EPRs with the same nominal rate. This happens because Gaussian ellipsoids for these variables tend to underestimate  uncertainty for higher nominal coverage rates and overestimate it for lower nominal coverage rates.
%\begin{figure}[t!]
%
%\begin{subfigure}{.5\textwidth}
%  \centering
%  \includegraphics[width=.7\linewidth]{OPEsPrice2}
%%  \caption{Price: Prediction ellipsoids of dimension 24}
%  \label{fig:OPESPrice}
%\end{subfigure}
%\vspace{+0.4em}
%\\
%\vspace{+0.4em}
%\begin{subfigure}{.5\textwidth}
%  \centering
%  \includegraphics[width=.7\linewidth]{OPEsWind2}
% % \caption{Wind Power: Prediction ellipsoids of dimension 24}
%  \label{fig:OPEsWind}
%\end{subfigure}
%\\
%\vspace{+0.4em}
%\begin{subfigure}{.5\textwidth}
%  \centering
%  \includegraphics[width=.7\linewidth]{OPEsPV2}
% % \caption{PV Power: Prediction ellipsoids of dimension 11}
%  \label{fig:OPEsPV}
%\end{subfigure}%
%
%\caption{19 Optimal EPRs with coverage ranging from 0.05 to 0.95 by 0.05 increments (from the lightest to the darkest), for three randomly selected days from the evaluation set for price, wind and PV power data. Character $ t $ denotes the day number in the evaluation dataset. \label{Fig:19OPEs}}
%\vspace{-0.5em}
%\end{figure}

In Figs. \ref{Fig:PEComparePV} and \ref{Fig:PEComparPrice2}, 95\% and 70\% EPRs  along with robust ellipsoids with budget uncertainty $ \varUpsilon=D^2=4 $, and 95\% and 70\% Gaussian ellipsoids for PV power data and price are depicted.  By comparing the EPRs in Figs.  \ref{Fig:19OPEs}, \ref{Fig:PEComparePV}, it is clear that the shape, rotation and the ratio of semi-major to semi-minor axes of the ellipsoids vary depending on the stochastic process of interest as well as the underling non-stationery uncertainty level in various days. As one can  notice, the sizes of price EPRs for different days are very close while the those of PV EPRs vary in various days. This happens because the covariance of electricity price forecasting errors changes very slowly in time while the rate is much faster for  PV power.  The robust ellipsoids with $ \varUpsilon=4 $ give the observed coverage rates of 90\%, 82\% and 82\% for price, wind and PV power, respectively. 

For the robust ellipsoids, as the dimension of multivariate random variable increases, the ratio of their volume to those of EPRs increases. For bivariate case as $ D^2=4 $ is small and close to $ \chi^{2}_{2}(0.95)= 5.6 $, the differences between the area of the ellipsoids are not very large. However, for $  D=24 $, $ \chi^{2}_{24}(0.95)= 36.4 $ comparing to $ {24}^2=576 $, the volume can be substantially larger.
\vspace{-0.6em}

\subsection{Skill verification and evaluation}
 To get a sense of the values of optimized scale parameters, Fig. \ref{fig:gamma} is provided. In this figure, the optimized $ \Upsilon $ for dimensions 24, 11 and 2 are shown. The maximum $ \varUpsilon^\alpha $ for  dimension 24 is about 70 which is much lower than $ 24^2$. 

 Comparing $ \varUpsilon $ for various coverage rates shows  as it is  expected, the ellipsoids with higher nominal coverage rates are more conservative and larger than those with lower nominal coverage rates.  Based on  \eqref{Vol_ellipsoid}, with the same centre and covariance matrix, the value of $ \Upsilon$ intensely impacts on the volume of a multidimensional ellipsoid. 

In order to evaluate the robustness and reliability of the EPRs, Fig. \ref{fig:QQ} is provided. In this figure, calibration of the EPRs along with that of the Gaussian ellipsoids obtained for wind power and price  of dimension 24 and PV power of dimension 11 (from 7 am to 5 pm) are illustrated.
%\begin{figure}[t!]
%\begin{subfigure}{.5\textwidth}
%  \centering
%  \includegraphics[width=.7\linewidth]{QQPrice2}
%  \caption{Price: Prediction ellipsoids of dimension 24}
%  \label{fig:QQPrice}
%\end{subfigure}
%\vspace{-0.1em}
%\\
%\vspace{-0.1em}
%\begin{subfigure}{.5\textwidth}
%  \centering
%  \includegraphics[width=.7\linewidth]{QQWind2}
%  \caption{Wind Power: Prediction ellipsoids of dimension 24}
%  \label{fig:QQWind}
%\end{subfigure}
%\vspace{-0.1em}
%\\
%\vspace{-0.1em}
%\begin{subfigure}{.5\textwidth}
%  \centering
%  \includegraphics[width=.7\linewidth]{QQPVDm11_2}
%  \caption{PV Power: Prediction ellipsoids of dimension 11}
%  \label{fig:QQPV}
%  \vspace{-0.6em}
%\end{subfigure}%
%
%\caption{Calibration of the prediction ellipsoids with nominal coverage ranging from 0.05 to 0.95 by 0.05 increments. \label{fig:QQ}}
%\vspace{-0.5em}
%\end{figure}

The artificial Gaussian data in Fig. \ref{fig:QQ} are  generated by  random draw from multivariate Gaussian distributions with $\mu_t$ and $\varSigma_t$, $ \forall t $. The scale parameters of the predicted ellipsoids for this data are  $ \chi^{2}_{D}(\alpha_i) $, $ \forall \alpha_i $. The highly calibrated ellipsoids fitted to the artificial Gaussian data in Fig. \ref{fig:QQ} reveals that if the multivariate random process is normally distributed, the prediction ellipsoids can  be fully characterized by \eqref{Eq:Chi-sq_Main}.

Looking at  Fig. \ref{fig:QQ}, one can observe that EPRs present close to ideal calibration by making a  comprise between robustness (calibration) and performance (conservativeness) based on \eqref{Eq:objective}. The Gaussian ellipsoids tend to overestimate joint uncertainty for low nominal coverage rates but underestimate it for higher nominal coverage rates. This also can be inferred from  Fig. \ref{fig:gamma}, where the scale parameters given by $ \chi^2_D(\alpha) $ for  Gaussian ellipsoids with lower coverage rates are much larger than those of EPRs with the same coverage rate while the relationship is opposite for the higher coverage rates. From Fig. \ref{fig:QQ}, one can easily  perceive very low robustness of the Gaussian prediction ellipsoids. The maximum deviations as high as 47\% makes the Gaussian ellipsoids very unreliable to be used in decision-making as they lead to biased analyses. On the other hand, EPRs generated for all three datasets offer  reasonably high calibration. 
%It is to be noted that while in~\cite{li2015modeling} the aim is to find the ellipsoid with the highest probability,  the highest probability achieved can be as low as 42\% for 24 hour look-ahead time.

In Table \ref{Table:DeviationScore}, the calculated skill scores are  given for three dimensions. Much better skill scores of EPRs  with respect to the Gaussian ellipsoids confirm higher predictive performance of proposed ellipsoids in terms of both conservativeness and probability guarantees.
\begin{table}[!t]
	\centering
	\caption{Scores of EPRs and Gaussian prediction ellipsoids}
	\label{Table:DeviationScore}
	\centering
	%\resizebox{\columnwidth}{!}{%
	\begin{tabular}{c|c|ccc}
		\specialrule{1.5pt}{0pt}{0pt}
		&                   & \textbf{D=2} & \textbf{D=11} & \textbf{D=24} \\ \specialrule{1pt}{0pt}{0pt}
		\multirow{2}{*}{\textbf{Price}} & \textbf{EPRs}     & 0.119       & 0.085         & 0.069         \\ \cline{2-5} 
		& \textbf{Gaussian} & 0.914        & 1.034         & 0.794         \\ \hline
		\multirow{2}{*}{\textbf{Wind}}  & \textbf{EPRs}     & 1.023        & 1.369         & 2.119         \\ \cline{2-5} 
		& \textbf{Gaussian} & 1.483        & 4.228         & 4.223         \\ \hline
		\multirow{2}{*}{\textbf{PV}}    & \textbf{EPRs}     & 0.570        & 0.562         & -             \\ \cline{2-5} 
		& \textbf{Gaussian} & 1.745        & 2.411         & -             \\ \specialrule{1.5pt}{0pt}{0pt}
	\end{tabular}
	%}
	\vspace{-0.3em}
\end{table}

To examine the efficiency of the proposed volume estimation method explained in Appendix \ref{Section:Volume}, the reader is referred to  Appendix \ref{Simulation Vol}.
\vspace{-0.6em} 

\vspace{-0.6em}
\section{Conclusion} \label{Conclusion}
 We  propose a generic approach to construct EPRs  with predefined probability levels and optimal conservativeness. These multivariate ellipsoidal uncertainty sets provide essential information for the problems which are temporally  coupled. In order to verify the applicability of the proposed method in characterizing multivariate uncertainty information for the stochastic processes with different underling stochasticity, three different datasets including data for wind and PV power and electrify price are deployed. It is proposed to use exponential smoothing method to estimate the covariance matrix for those stochastic processes like electricity price with either constant or slow-moving covariance matrix. DCC-GARCH model is preferred for those stochastic processes like wind and PV with time-varying covariance. The simulation results showed that for all three case studies, Gaussian ellipsoids do not characterize the inherent joint uncertainty. They present very low robustness as a result of either overestimation or underestimation of the uncertainty level. This work provides a comprehensive framework for both generation and evaluation of ellipsoidal uncertainty sets  which have been used in robust optimization. The proposed scheme is able to track and predict the existing uncertainty level in time and characterize the EPRs such that they provide the desired probability level with optimal volume.  The proposed skill score is used to evaluate the predictive performance of the EPRs. The results confirm that the proposed approach is able to generate EPRs with acceptable reliability and conservativeness. The proposed framework can be applied to variety of the decisions-making problems which involve correlated random variables.

\vspace{-0.2em}
\appendices
\section{Estimating volume of EPRs for bounded variables}
%\section{Volume of Ellipsoids with Bound Constraint}
\label{Section:Volume}
The idea of estimating volume of EPRs is to generate $ N $ random samples in the feasible range and then calculate the proportion of those points which lie in the ellipsoids. If the limits in all dimensions are the same, then the feasible range forms  a hyper-cube. The estimated volume of the part of the ellipsoid inscribed in the feasible hyper-cube ($V^e$)  is considered to be
 \begin{equation}
 \label{VolumeGeneral}
V^e={N' V^c}/{N}
 \end{equation}
with $ N' $ as the number of $ D $-dimensional points enveloped by the ellipsoid and $ V^c $ is the volume of the bounded hyper-cube. 

\begin{figure}[!t]
	\vspace{-0.4em}
	%\begin{center}
	\centering
	\includegraphics[width=8cm,height=8cm]{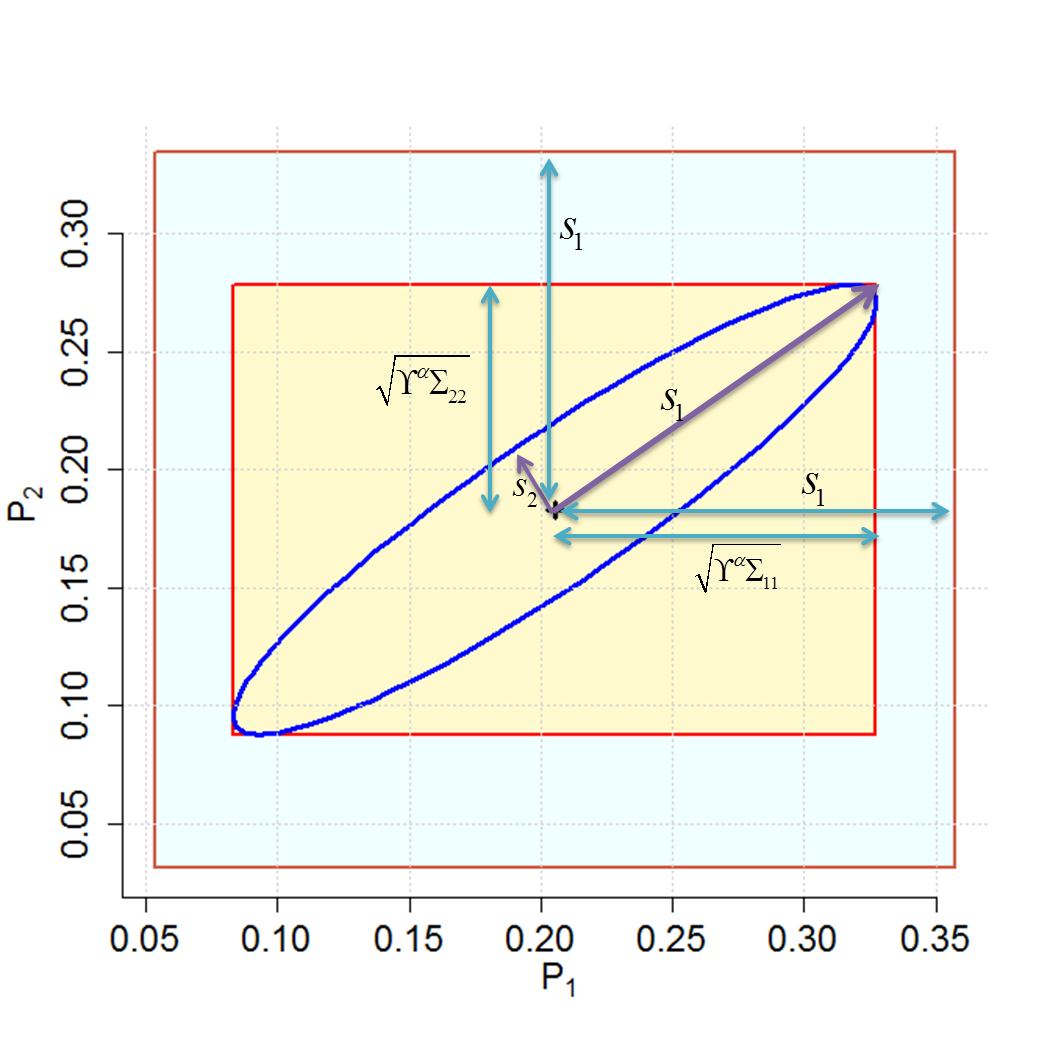}
	\caption{A typical prediction ellipsoid of dimension two  along with a hypercube and a hyper-rectangular enclosed it.}
	\label{Fig:Semiaxis}
	%\end{center}
	\vspace{-0.4em}
\end{figure}

The Monte Carlo method converges very slowly and it requires a large $ N $ to allow for a reasonable estimation. To increase computational efficiency, one can generate the random samples from a smaller geometry enclosing the ellipsoid. In~\cite{li2015modeling}, it is suggested to generate samples from a hyper-cube with edges of  equal to twice the largest semi-axis of the ellipsoid as
 \begin{equation}
L=2 \max \{\lambda_i^{-1/2 } \quad i=1,...,D\} 
 \end{equation}
with $ \lambda_i, \quad i=1,...,D $ as the eigenvalues of $ \varSigma^{-1}/\varUpsilon^{\alpha} $. In this case, $ V^c $ in \eqref{VolumeGeneral} is $ (L)^D $. The idea is illustrated in Fig. \ref{Fig:Semiaxis} for an ellipse with $ S_1 $ and $ S_2 $ as its semi-minor and semi-major axes. Although generating samples in $ (L)^D $ hyper-cube reduce the computational burden, still the method is computationally extensive specially in higher dimensions. The Monte Carlo method will be improved if the samples are generated from the smallest hyper-rectangular circumscribed the ellipsoid. To find the minimum volume hyper-rectangular, it is sufficient to find the  outermost point for each coordinate of the ellipsoid. 

If $\textbf{x}$ is located on the ellipsoidal surface, it satisfies $(x - \mu)^t\Sigma^{-1}(x-\mu) = \varUpsilon^{\alpha}$. For $\textbf{x}$, outward normal is pointing towards the direction
 \begin{equation}
 	\vspace{-0.2em}
\nabla \left[(\textbf{x} - \mu)^t\Sigma^{-1}(\textbf{x}-\mu) - \varUpsilon^{\alpha} \right] \propto \Sigma^{-1}
(x-\mu)
	\vspace{-0.2em}
 \end{equation}
 In order  to maximize $\,x\cdot n = n^t x\,$ along  direction $n$, it is required to have
 \begin{equation}
 	\vspace{-0.4em}
\Sigma^{-1} (\textbf{x}-\mu) \propto n\quad\iff\quad \textbf{x} - \mu = \kappa \Sigma n\quad\text{ for some }\kappa > 0
 \end{equation}
Substituting $ \kappa \Sigma n $ into the equation of the ellipsoidal surface gives
 \begin{multline}
\varUpsilon^{\alpha} = (\textbf{x}-\mu)^t \Sigma^{-1} (\textbf{x}-\mu) = \kappa^2 n^t \Sigma \Sigma^{-1}\Sigma n = \kappa^2 n^t\Sigma n 
\\
 \implies\quad\kappa^2 = \frac{\varUpsilon^{\alpha}}{n^t \Sigma n}
 \end{multline}
Therefore, the point which maximizes $ \textbf{x}.n $ is given by
\begin{equation}
\textbf{x}^{\max} = \mu + \sqrt{\frac{\varUpsilon^{\alpha}}{n^t \Sigma n}}\Sigma n
 \end{equation}
The extreme values of the ellipsoid in dimension $ i $ are given as 
\begin{equation}
{x}^{\max/\min}_{i} = \mu_i \pm \sqrt{\varUpsilon^{\alpha}\Sigma_{ii}}  \quad \forall i
\end{equation}
with $ \Sigma_{ii} $ as the $ i^{th} $ diagonal element of $ \Sigma $.

The length of each edge of the proposed minimum-volume hyper-rectangular is $ L_i= {x}^{\max}_{i}-{x}^{\min}_{i} $ and the $ V^c $ can be calculated by multiplication of the length  of $ D $ sides of the hyper-rectangular. In Fig. \ref{Fig:Semiaxis}, the proposed hyper-rectangular is depicted for a typical EPR. The difference between volume of the hyper-cube proposed in~\cite{li2015modeling} and the hyper-rectangular proposed here can be large in higher dimensions.
%The convergence of this method is too slow to be used in the optimization process because $ N $ should be sufficiently large to allow for a reasonable estimation. However, if one needs to calculate the feasible volume of prediction ellipsoids for other purposes, the Monte carlo method should be conducted. In Table \ref{Table:Volume Compare}, because all the ellipsoids for the same day and the same data  are characterized by the same center and covariance matrix and they differ  only on the scale parameters, the comparisons are still true unless the ellipsoid in all dimensions exceed the feasible limits. In that case, the feasible volume can be considered equal to the volume of the same dimensional hypercube for all the corresponding ellipsoids. It is to be noted that the feasible $ D $-dimensional geometrical region can be a hyper-rectangle if the feasible limits for each dimension is different. For example, for the case of PV generation, if power is considered to be limited to the clear sky power, the feasible limits for each hour is different from others.
\vspace{-0.5em}
\section{Simulation Results of volume estimation for bounded random variables}
\label{Simulation Vol}
 \vspace{-0.1em}
Wind and PV power both are double-bounded random variables between zero and 1 pu. Electricity price also can have upper and lower limits depending on power market regulations. As shown in Fig. \ref{Fig:19OPEs},  there is a chance that EPRs exceed the practical limits. Here, as explained in Appendix \ref{Section:Volume},  the volume of intersection of the EPRs and the  polyhedrons are estimated using Monte Carlo method. By following~\cite{li2015modeling}, a metric named Absolute Deviation Percentage ($ADP$) is used to examine the efficiency of volume estimation. $ADP$ is given by
\begin{equation}
ADP =\frac{|V^c_N -V^c_{N+\bigtriangleup N}|}{V^c_N}
\end{equation}
with $V^c_N$ as the estimated volume with $N$ random samples and $\bigtriangleup N=N/100$. $ADP$ values along with system simulation time for five sets of random samples are given in Table \ref{Table:ADP}.  In~\cite{li2015modeling}, $N=1200$, $N=20,000$ and $N=500,000$ resulted in $ADP$ 0.952, 0.322 and 0.116, respectively. Comparing these values with those reported in Table \ref{Table:ADP},  one can conclude that the smaller circumscribed polyhedron proposed in Appendix \ref{Section:Volume} leads to more efficient estimation of volumes of EPRs. The results in Table \ref{Table:ADP} are based on estimation of 90\% EPRs. 
\begin{table}[!t]
	\centering
	\caption{Absolute deviation percentage of estimated volume of ellipsoids}
	\label{Table:ADP}
	\centering
	\resizebox{\columnwidth}{!}{%
		\begin{tabular}{c|ccccc}
			\specialrule{1.5pt}{0pt}{0pt}
			& \textbf{N=5,000} & \textbf{N=10,000} & \textbf{N=20,000} & \textbf{N=50,000} & \textbf{N=100,000} \\ \specialrule{1pt}{0pt}{0pt}
			\textbf{ADP (\%)} & 0.614           & 0.550         & 0.297           & 0.207    &  0.095     \\ \hline
			\textbf{Simulation Time (s)} & 0.0008           & 0.0028           & 0.0094           & 0.0079       & 0.0174    \\ \specialrule{1.5pt}{0pt}{0pt}
		\end{tabular}
	}
\vspace{-0.3em}
\end{table}
%Some text for the appendix.

% use section* for acknowledgement

%\begin{thebibliography}{9}
	%\bibliography{reftest}
	
%\end{thebibliography}

%\section*{Acknowledgment}
%Hello~\cite{article123}
% % % % % % % % % %
\vspace{-0.3em}
\bibliographystyle{IEEEtran}
\vspace{-0.34em}
\bibliography{reftest}

% Generated by IEEEtran.bst, version: 1.12 (2007/01/11)
\begin{thebibliography}{10}
\providecommand{\url}[1]{#1}
\csname url@samestyle\endcsname
\providecommand{\newblock}{\relax}
\providecommand{\bibinfo}[2]{#2}
\providecommand{\BIBentrySTDinterwordspacing}{\spaceskip=0pt\relax}
\providecommand{\BIBentryALTinterwordstretchfactor}{4}
\providecommand{\BIBentryALTinterwordspacing}{\spaceskip=\fontdimen2\font plus
\BIBentryALTinterwordstretchfactor\fontdimen3\font minus
  \fontdimen4\font\relax}
\providecommand{\BIBforeignlanguage}[2]{{%
\expandafter\ifx\csname l@#1\endcsname\relax
\typeout{** WARNING: IEEEtran.bst: No hyphenation pattern has been}%
\typeout{** loaded for the language `#1'. Using the pattern for}%
\typeout{** the default language instead.}%
\else
\language=\csname l@#1\endcsname
\fi
#2}}
\providecommand{\BIBdecl}{\relax}
\BIBdecl

\bibitem{morales2013integrating}
J.~M. Morales, A.~J. Conejo, H.~Madsen, P.~Pinson, and M.~Zugno,
  \emph{Integrating renewables in electricity markets: Operational
  problems}.\hskip 1em plus 0.5em minus 0.4em\relax Springer Science \&
  Business Media, 2013, vol. 205.

\bibitem{zhang2014review}
Y.~Zhang, J.~Wang, and X.~Wang, ``Review on probabilistic forecasting of wind
  power generation,'' \emph{Renewable and Sustainable Energy Reviews}, vol.~32,
  pp. 255--270, 2014.

\bibitem{tastu2013short}
J.~Tastu, ``Short-term wind power forecasting: probabilistic and space-time
  aspects,'' \emph{PhD thesis}, 2013.

\bibitem{zhang2011chance}
H.~Zhang and P.~Li, ``Chance constrained programming for optimal power flow
  under uncertainty,'' \emph{IEEE Transactions on Power Systems}, vol.~26,
  no.~4, pp. 2417--2424, 2011.

\bibitem{sousa2011robust}
A.~A. Sousa, G.~L. Torres, and C.~A. Canizares, ``Robust optimal power flow
  solution using trust region and interior-point methods,'' \emph{IEEE
  Transactions on Power Systems}, vol.~26, no.~2, pp. 487--499, 2011.

\bibitem{jabr2013adjustable}
R.~A. Jabr, ``Adjustable robust opf with renewable energy sources,'' \emph{IEEE
  Transactions on Power Systems}, vol.~28, no.~4, pp. 4742--4751, 2013.

\bibitem{wu2012comparison}
L.~Wu, M.~Shahidehpour, and Z.~Li, ``Comparison of scenario-based and interval
  optimization approaches to stochastic scuc,'' \emph{IEEE Transactions on
  Power Systems}, vol.~27, no.~2, pp. 913--921, 2012.

\bibitem{bessa2015marginal}
R.~J. Bessa, ``From marginal to simultaneous prediction intervals of wind
  power,'' in \emph{Intelligent System Application to Power Systems (ISAP),
  2015 18th International Conference on}.\hskip 1em plus 0.5em minus
  0.4em\relax IEEE, 2015, pp. 1--6.

\bibitem{kolsrud2007time}
D.~Kolsrud, ``Time-simultaneous prediction band for a time series,''
  \emph{Journal of Forecasting}, vol.~26, no.~3, pp. 171--188, 2007.

\bibitem{li2011simultaneous}
J.~S.-H. Li and W.-S. Chan, ``Simultaneous prediction intervals: An application
  to forecasting us and canadian mortality,'' 2011.

\bibitem{chassein2016min}
A.~Chassein and M.~Goerigk, ``Min-max regret problems with ellipsoidal
  uncertainty sets,'' \emph{arXiv preprint arXiv:1606.01180}, 2016.

\bibitem{bertsimas2004robust}
D.~Bertsimas and M.~Sim, ``Robust discrete optimization under ellipsoidal
  uncertainty sets,'' 2004.

\bibitem{guan2014uncertainty}
Y.~Guan and J.~Wang, ``Uncertainty sets for robust unit commitment,''
  \emph{IEEE Transactions on Power Systems}, vol.~3, no.~29, pp. 1439--1440,
  2014.

\bibitem{li2015modeling}
P.~Li, X.~Guan, J.~Wu, and X.~Zhou, ``Modeling dynamic spatial correlations of
  geographically distributed wind farms and constructing ellipsoidal
  uncertainty sets for optimization-based generation scheduling,'' \emph{IEEE
  Transactions on Sustainable Energy}, vol.~6, no.~4, pp. 1594--1605, 2015.

\bibitem{engle2002dynamic}
R.~Engle, ``Dynamic conditional correlation: A simple class of multivariate
  generalized autoregressive conditional heteroskedasticity models,''
  \emph{Journal of Business \& Economic Statistics}, vol.~20, no.~3, pp.
  339--350, 2002.

\bibitem{pope2008algorithms}
S.~B. Pope, ``Algorithms for ellipsoids,'' \emph{Cornell University Report No.
  FDA}, pp. 08--01, 2008.

\bibitem{zakamulin2015test}
V.~Zakamulin, ``A test of covariance-matrix forecasting methods,'' \emph{The
  Journal of Portfolio Management}, vol.~41, no.~3, pp. 97--108, 2015.

\bibitem{engle2001theoretical}
R.~F. Engle and K.~Sheppard, ``Theoretical and empirical properties of dynamic
  conditional correlation multivariate garch,'' National Bureau of Economic
  Research, Tech. Rep., 2001.

\bibitem{website55}
\BIBentryALTinterwordspacing
Global energy forecasting competition 2014 probabilistic solar power
  forecasting. [Online]. Available: \url{https://crowdanalytix.com}
\BIBentrySTDinterwordspacing

\bibitem{hong2016probabilistic}
T.~Hong, P.~Pinson, S.~Fan, H.~Zareipour, A.~Troccoli, and R.~J. Hyndman,
  ``Probabilistic energy forecasting: Global energy forecasting competition
  2014 and beyond,'' \emph{International Journal of Forecasting}, vol.~32,
  no.~3, pp. 896--913, 2016.

\bibitem{huang2016semi}
J.~Huang and M.~Perry, ``A semi-empirical approach using gradient boosting and
  k-nearest neighbors regression for gefcom2014 probabilistic solar power
  forecasting,'' \emph{International Journal of Forecasting}, vol.~32, no.~3,
  pp. 1081--1086, 2016.

\bibitem{golestaneh2016very}
F.~Golestaneh, P.~Pinson, and H.~B. Gooi, ``Very short-term nonparametric
  probabilistic forecasting of renewable energy generation—with application
  to solar energy,'' \emph{IEEE Transactions on Power Systems}, vol.~31, no.~5,
  pp. 3850--3863, 2016.

\end{thebibliography}
\end{document}